\documentclass[amsmath,amssymb,twocolumn]{revtex4}
\usepackage{graphicx}
\usepackage{amssymb}
\usepackage{bm}
\usepackage{color} 
\usepackage{hyperref}

\begin{document}

\title{Analysis of the near-resonant fluorescence spectra of a single rubidium atom localized in a three-dimensional optical lattice}
\author{Wookrae Kim}
\author{Changwon Park}
\author{Jung-Ryul Kim}
\author{Yea-Lee Lee}
\author{Jisoon Ihm}
\author{Kyungwon An}
\email{Kwan@phya.snu.ac.kr}
\affiliation{Department of Physics and Astronomy, Seoul National University, Seoul 151-747, Korea}
\date{\today}
\begin{abstract}
Supplementary information is presented on the recent work by W.\ Kim {\em et al.} on the matter-wave-tunneling-induced broadening in the near-resonant spectra of a single rubidium atom localized in a three-dimensional optical lattice in a strong Lamb-Dicke regime. 
\end{abstract}

\maketitle

\section{Introduction}
The purpose of this brief article is to provide the supplementary information on the recent work by W.\ Kim {\em et al.} \cite{W-Kim2010} on the near-resonant spectra of a single $^{85}$Rb atom localized in a three-dimensional (3D) optical lattice. This article is organized as follows. In Sec.\,2 the formula for the 3D optical lattice in a phase-stabilized magneto-optical trap (MOT) is derived. In Sec.\,3 the data analysis details for the spectral lineshape fitting are given. In Sec.\,4 supplementary data on many-atom spectra with a fixed depletion rate to the inaccessible $F=2$ hyperfine level are presented.

\section{Three-dimensional optical lattice formed in a phase-stabilized MOT}

\begin{figure}[t]
\includegraphics[width=3.4in]{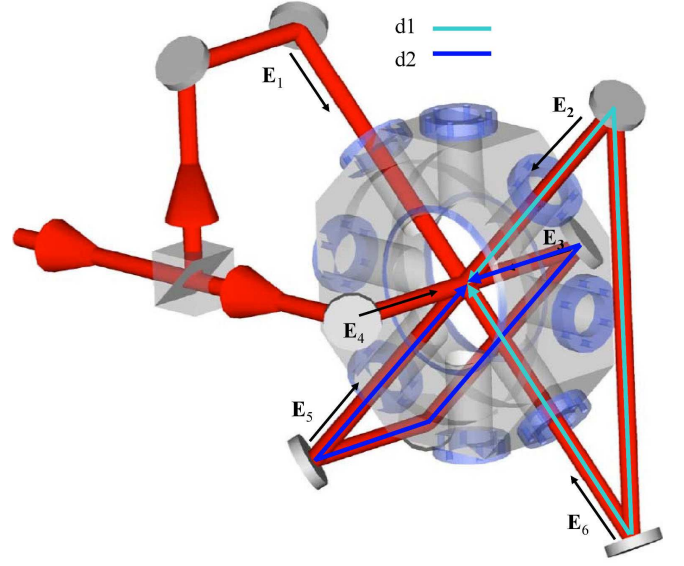}
\caption{Six beam configuration for a time-phase stabilized lattice structure.}
\label{fig1}
\end{figure}

A three-dimensional optical lattice can be established in a phase-stabilized MOT. Phase stabilization can be done by passive means, {\em i.e.}, by making a standing wave formed by a trap laser intersect with itself at right angle twice by folding it \cite{time-phase compensation} as shown in Fig.\,\ref{fig1}. In this configuration, the electric fields of the trap laser can be expressed as follows, assuming a unit amplitude for each field,
\begin{eqnarray}
\mathbf{E}_1&=&\cos(\omega t-kx+\phi _1)\mathbf{\hat{j}}+\sin(\omega t-kx+\phi_1)\mathbf{\hat{k}}, \nonumber\\
\mathbf{E}_2&=&\cos(\omega t-ky+\phi _2)\mathbf{\hat{k}}+\sin(\omega t-ky+\phi _2)\mathbf{\hat{i}}, \nonumber\\
\mathbf{E}_3&=&\cos(\omega t+kz+\phi _3)\mathbf{\hat{i}}+\sin(\omega t+kz+\phi _3)\mathbf{\hat{j}}, \nonumber\\
\mathbf{E}_4&=&\cos(\omega t-kz+\phi _4)\mathbf{\hat{i}}-\sin(\omega t-kz+\phi _4)\mathbf{\hat{j}}, \nonumber\\
\mathbf{E}_5&=&\cos(\omega t+ky+\phi _5)\mathbf{\hat{k}}-\sin(\omega t+ky+\phi _5)\mathbf{\hat{i}}, \nonumber\\
\mathbf{E}_6&=&\cos(\omega t+kx+\phi _6)\mathbf{\hat{j}}-\sin(\omega t+kx+\phi _6)\mathbf{\hat{k}},\nonumber
\end{eqnarray}
where $\mathbf{\hat{i}}$, $\mathbf{\hat{j}}$ and $\mathbf{\hat{k}}$ are unit vectors parallel to the propagation directions of $\mathbf{E}_1$, $\mathbf{E}_2$ and $\mathbf{E}_4$, respectively, $k=\omega/c$, and $\phi_1, \phi_2, \ldots, \phi_6 $ are the phase factors for $\mathbf{E}_1, \mathbf{E}_2 \ldots, \mathbf{E}_6$, respectively.
Since each beam path is not independent from the others, there exists some constraints on the phase factors. 
The mathematical form for the total electric field in the intersection region is given by \cite{time-phase compensation}. 
\begin{eqnarray}
\mathbf{E}_{\rm tot}&=& 2\cos\omega \tau \left[(\cos kz -\sin kY)\mathbf{\hat{i}}+(\cos kX+\sin kz)\mathbf{\hat{j}} \right.\nonumber \\
&+&\left.(\cos kY-\sin kX)\mathbf{\hat{k}}\right], \nonumber
\end{eqnarray}
where
\begin{eqnarray}
\tau = t+\frac{\phi_1+\phi_6}{2\omega} &=& t+\frac{\phi_2+\phi_5}
{2\omega} = t+\frac{\phi_3+\phi_4}{2\omega},\nonumber\\
 X &=& x+d_1 + d_2, \nonumber\\ 
 Y &=& y + d_2, \nonumber
\end{eqnarray}
where $d_1$ and $d_2$ are the path lengths depicted as thick colored lines in Fig.\,\ref{fig1}.
Note that the total time phase $\omega \tau$ term is common, thus the interference pattern is insensitive to time-phase jittering although the whole pattern may slowly shift due to the fluctuating phase factors or path lengths $d_1$ and $d_2$.
The light shift potential due to interference pattern of the standing waves, corresponding to the optical-lattice potential structure in Fig.\,1(a) of Ref.\,\cite{W-Kim2010}, can then be calculated as
\begin{eqnarray}
V\propto - \left | \mathbf{E}_{\rm tot} \right |^{2}
  &=& -2 (3-2\sin kY \cos kz +2\sin kX \sin kz\nonumber \\
  &+&2\cos kX \cos kY).
\end{eqnarray}

\section{How to fit the single-atom fluorescence spectra to obtain the fit curve in Fig.\,4 of Ref.\,\cite{W-Kim2010}}

In Ref.\,\cite{W-Kim2010} we trapped a single rubidium atom in the optical lattice in a phase-stabilized MOT and measured its near-resonant spectra for various micro-potential depths of the optical lattice. The full-width-at-half-maximum of the central Rayleigh peak was then plotted as a function of the micro-potential depth in Fig.\,4 there.

The fit curve in Fig.\,4 of Ref.\,\cite{W-Kim2010} was obtained as follows. For a given oscillation frequency $\nu_{\rm osc}$ of the micropotential, the observed lineshape of the Rayleigh peak is fit by the following convolution integrals.
\begin{equation}
{\cal L}_{\rm Rayleigh}(\omega)=P_g S_g {\cal L}_g (\omega)+P_e S_e {\cal L}_e(\omega)
\end{equation}
where $P_g(P_e)$ is the population in the ground (excited) vibrational level including a degeneracy factor (1 for the ground and 2 for the excited level) and $S_g (S_e)$ is the Rayleigh scattering cross section of the ground (excited) vibrational levels. Lineshape ${\cal L}_g$ is given by a Voigt integral (with respect to a resonance frequency):
\begin{equation}
{\cal L}_g(\omega)=\int d\omega' L(\omega'; \gamma_g) G(\omega-\omega'; \sigma_g)
\end{equation}
where $L(\omega; \gamma)$ is a Lorentzian given by
\begin{equation}
L(\omega; \gamma)=\frac{1}{\pi}\frac{\gamma}{\omega^2+\gamma^2},
\end{equation}
and $G(\omega;\sigma)$ is a Gaussian given by
\begin{equation}
G(\omega;\sigma)=\frac{1}{\sigma\sqrt{2\pi}}\exp{\left(-\frac{\omega^2}{2\sigma^2}\right)}.
\end{equation}
Parameter $\gamma_g$ is given by a sum of $\gamma_{\rm tun}^{(g)}$ the $s$-state bandwidth due to tunneling (given by the band calculation of Ref.\,\cite{W-Kim2010}) and $\gamma_{\rm dep}$ the depletion-induced broadening of the excited electronic state (independently measured). Parameter $\sigma_g$ is given by the square root of the sum of squares of $\sigma_{\rm res}$, the spectral resolution of our system (given by experiment), and $\sigma_{\rm inh}^{(g)}$, a small inhomogeneous broadening, which is a fitting parameter. Lineshape ${\cal L}_e$ is similarly defined with $\gamma_e=\gamma_{\rm tun}^{(e)}+\gamma_{\rm dep}$ and $\sigma_e^2=\sigma_{\rm res}^2+(\sigma_{\rm inh}^{(e)})^2$, where $\gamma_{\rm tun}^{(e)}$ is the $p$-state bandwidth due to tunneling and $\sigma_{\rm inh}^{(e)}$, another fitting parameter, is the inhomogeneous broadening associated with the excited vibrational level. 

\begin{figure}[t]
\includegraphics[width=3.4in]{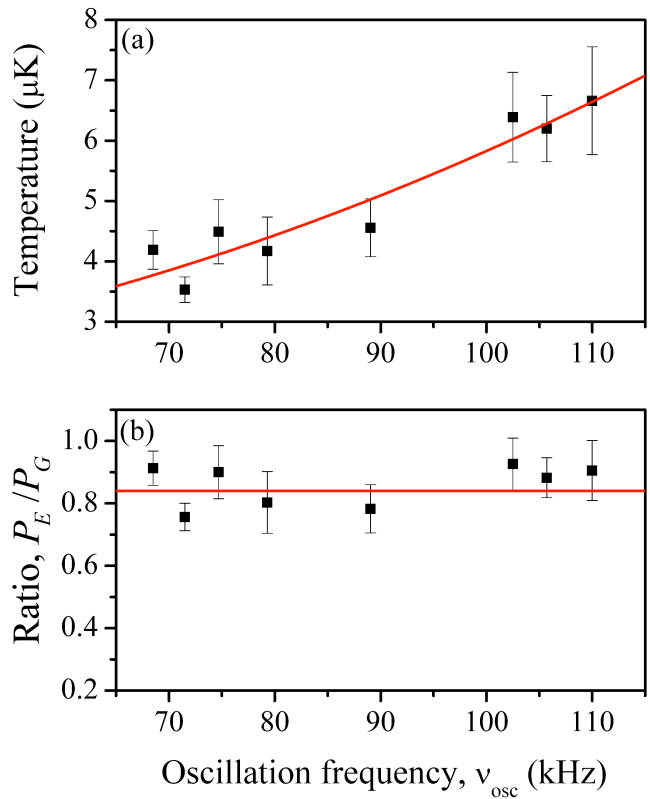}
\caption{(a) Observed temperature of the trapped atoms versus the oscillation frequency. Solid line is a quadratic fit. (b) The resulting population ratio $P_E/P_G$ versus the oscillation frequency.}  
\label{fig3}
\end{figure}

The populations are determined by the temperature of the atoms \cite{population}. The temperature increases as the potential depth increases because the atomic momentum diffusion is proportional to the trap laser power \cite{Dalibard-JOSA89, Gatzke-PRA97}. We have confirmed this relation in our experiment. For the range of trap depth covered in our experiment, the temperature varied from 3 $\mu$K to 6 $\mu$K as shown in Fig.\,\ref{fig3}(a). However, the Boltzmann factor $\exp(-h\nu_{\rm osc}/k_BT)$, which determines the population ratio $P_E/P_G$, remains almost constant, so $P_E/P_G\simeq 0.84$ as shown in Fig.\,\ref{fig3}(b). Since the Rayleigh scattering cross section is proportional to $n+1/2$ for $n$th vibrational level, $S_E/S_G=3$ \cite{Grynberg}, we then obtain $P_E S_E: P_G S_G=0.72:0.28$.

There are two fitting parameters $\sigma_{\rm inh}^{(g)}$ and $\sigma_{\rm inh}^{(e)}$. However, we obtain $\sigma_{\rm inh}^{(g)}\simeq \sigma_{\rm inh}^{(e)}\simeq (1.8\pm 0.1)$ kHz within the fitting uncertainty for all $\nu_{\rm osc}$ in Fig.\,4 of Ref.\,\cite{W-Kim2010}. The full-width-at-half-maximum of the resulting lineshape curve of Eq.\,(2) as a function of $\nu_{\rm osc}$ is our final fit curve in Fig.\,4 of Ref.\,\cite{W-Kim2010}. It well reproduces the trend of the Rayleigh-peak linewidth at small $\nu_{\rm osc}$.

\section{Rayleigh-peak linewidth data with a constant depletion rate into $F=2$ levels}
\begin{figure}[t]
\includegraphics[width=3.4in]{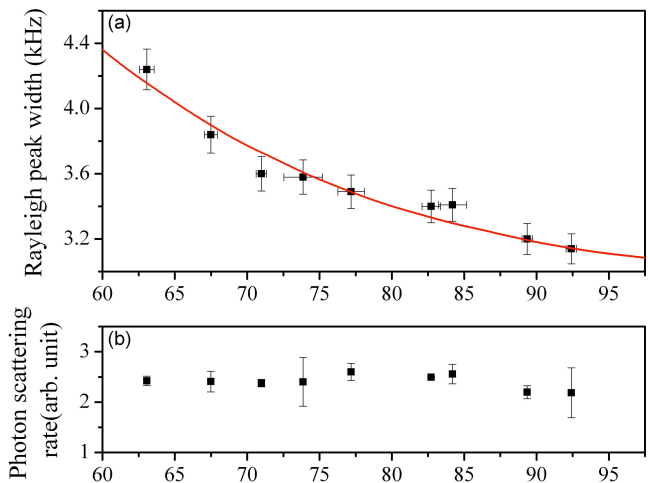}
\caption{Observed linewidth of the Rayleigh peak versus the oscillation frequency when the optical pumping (depletion) rate into inaccessible hyperfine levels is kept constant within 5.9\% deviation. In this case, the tunneling effect is responsible for the observed dependence of the linewidth as the oscillation frequency.}  
\label{fig2}
\end{figure}

In order to isolate the tunneling effect in the Rayleigh-peak linewidth, we have also measured the spectra for various trap depths or the oscillation frequency $\nu_{\rm osc}$ while keeping the depletion rate into the inaccessible $F=2$ hyperfine levels \cite{W-Kim2010} almost constant. Note that the depletion rate is proportional to the population of the excited electronic states whereas the potential depth is proportional the light shift of the ground electronic state. We carefully selected sets of MOT parameters producing the same depletion rate but different trap depths or different $\nu_{\rm osc}$.  The experimental results obtained with hundreds of atoms are shown in Fig.\,\ref{fig2}. 
As mentioned in Ref.\,\cite{W-Kim2010}, we could not perform the same experiment for a single atom for the given range of $\nu_{\rm osc}$ because the signal level became too low for small and large $\nu_{\rm osc}$.

In our analysis the observed Rayleigh peak was fit by Eq.\,(2), but this time with three fitting parameters: $\sigma_{\rm inh}^{(g)}$, $\sigma_{\rm inh}^{(e)}$ and $\gamma_{\rm dep}$. The parameter $\gamma_{\rm dep}$ is kept constant for all oscillation frequencies since the population of the excited electronic states were kept almost constant in our experiment as shown in Fig.\,\ref{fig2}(b).  We obtained $\gamma_{\rm dep}\simeq (1.3\pm 0.1)$ kHz and  $\sigma_{\rm inh}^{(g)}\simeq (1.7\pm 0.1)$ kHz as in the single-atom experiment, but  $\sigma_{\rm inh}^{(e)}\simeq (2.2\pm 0.1)$ kHz, 22\% larger than that in the single-atom case, indicating additional broadening due to many atoms. Nonetheless, it is clear that the linewidth decreases as the oscillation frequency increases in Fig.\,\ref{fig2}, confirming the dependence of the tunneling effect on the oscillation frequency.

\end{document}